\documentclass[11pt,dvips]{article}

\usepackage{epsfig,times} 
\usepackage{picinpar}
\usepackage{wrapfig}
\usepackage{floatflt}
\makeatletter
\renewcommand{\section}{\@startsection{section}{1}{0in}
	{0.4\baselineskip}{0.1\baselineskip}{\Large\bf}}
\renewcommand{\subsection}{\@startsection{subsection}{2}{0in}
	{0.25\baselineskip}{-\baselineskip}{\large\bf}}
\renewcommand{\subsubsection}{\@startsection{subsubsection}{3}{0in}
	{0.1\baselineskip}{-\baselineskip}{\normalsize\bf}}
\makeatother
\begin{document}

%
\makeatletter\newcommand{\ps@icrc}{
\renewcommand{\@oddhead}{\slshape{OG.9.9.09}\hfil}}
\makeatother\thispagestyle{icrc}
%
%

%
\begin{center}
{\Large \bf Core, Mantle, Crust: Imaging the Earth's
Interior with Ultra-High Energy Neutrino Tomography } 
\end{center}

\begin{center}
%
%
{ \bf  John P.
Ralston$^1$, Pankaj Jain$^2$,
 and
George M. Frichter$^3$}\\
 
{\it $^1$Department of Physics and Astronomy \\ University of Kansas,
Lawrence, KS 66045-2151, USA\\$^2$Physics Department, I.I.T., Kanpur, India
208016 \\$^3$Department of Physics, Florida State
University, Talahassee, FL 32306-3016 USA, \\ } 
\end{center}

\begin{center}
{\large \bf Abstract\\}
\end{center}
\vspace{-0.5ex}
%
%
 We report on the feasibility of using an isotropic flux of cosmic
neutrinos in the energy range of 10 to 10,000 TeV to study the
 interior geophysical structure of the Earth. The angular distribution
 of events in a $\sim {\rm km}^3$-scale neutrino telescope can be inverted
 to yield
 the Earth's nucleon density
distribution.  The approach is independent of previous seismic methods.
 The energy spectrum of
the neutrino primaries can also be determined from
consistency with the angular distribution. Depending on neutrino flux,
a reasonable density determination can be obtained with a
 few hundred receivers operating for a few years.
%

\vspace{1ex}

%
%
\section{Introduction:}
\label{intro.sec}
The character of the Earth's interior remains a fascinating subject.
The density profile
has been deduced only indirectly by methods with model-dependence. We
report on a novel way to create a direct `snapshot' of the nucleon
density from core to crust, by exploiting tomography with ultra-high
energy neutrinos of cosmic origin (Jain, Ralston and Frichter 1999).
We assume there exists a reasonably isotropic source of ultra-high
 energy (UHE) neutrinos from cosmological sources.
 The possible sources include unresolved active galactic
nuclei, gamma ray bursts,
secondary emissions from cosmic rays, {\it etc}.

Neutrino Tomography exploits the information in the
 observed {\it angular distribution}. After passing through
 the Earth, the surviving neutrino flux $\Phi_\nu$ measured at nadir
 angle $\theta$ is given by

\begin{equation} \Phi_{\rm surv}(E,\theta)=\Phi_\nu(E)e^{-\sigma_{\rm
eff}(E)R n(R) f(\theta)}\:, \label{eqtwo} \end{equation}
 where $R$ is the Earth's radius  
and the function $f(\theta)$ is proportional to the  
integrated nucleon density along the chord $0<z<2R{\rm cos}(\theta)$.
Using the absorption cross section $\sigma_{\rm eff}(E)$ from particle
 physics (Frichter {\it et al} 1995, Gandhi {\it et al} 1998),
there is enough information in $\Phi_{\rm surv}(E,\theta)$ to 
 invert an integral transform, yielding
 the nucleon density. Thus our approach differs
fundamentally from previous studies concentrating on exploiting point
 neutrino sources (Kuo {\it et al} 1995, and references therein).
 An advantage of our approach is
that the ``whole sky'' is used as a source
for  tomography. In addition,
the statistical cost of binning events periodically in time required
 for point sources
 is eliminated. As a result the results appear promising.

Event rates scale with the primary fluxes. Our method
is not overly dependent on flux estimates: The flux normalization is actually
 irrelevant, and only contributes to the statistical errors. We use fluxes
consistent with current estimates and their updated normalizations
 (Stecker {\it et al} 1992, Szabo and Protheroe 1994, Protheroe and Szabo 1992). 
Event rates also scale with detector size, and detection efficiency in the
relevant energy range.  We focus 
on future experiments with a detection volume of order 1 km$^3$ (KM3).
 We use Monte Carlo simulations of
 the RICE pilot experiment currently running in conjunction with AMANDA
at the South Pole. The RICE project (Frichter {\it et al} 1996a,
 Frichter {\it et al} 1996b)
 uses coherent 
radio emission from neutrino-induced electromagnetic showers,
 which is the most efficient known method for neutrino energies
of roughly 100 TeV and above.
 Radio detection is not mandatory, of course, and 
any method generating comparable statistics and angular pointing
accuracy should work as well.

The energy dependence of the primary neutrino flux is also determined by
the procedure.  This is unexpected and welcome. Energy determination occurs
because the interaction cross
section and detector efficiencies are energy dependent.  (Both the interaction
cross section and the detector efficiency rise like a power of energy. 
These tend to compensate the fall of flux with energy, yielding a 
comparatively flat, broad dominant region of 1-10 PeV primary neutrino
 energy.) To exploit this we first need a well determined angular
 distribution integrated over energy.
We also need initial data, which we realistically assume is poorly determined,
on the energy spectrum integrated over angle.
Our procedure then alternately iterates the energy spectrum and density
 to obtain
 consistency with the angular spectrum. At no point is the joint
 energy and angular distribution needed. 

 For a wide range
of trial density distributions the procedure converges
 within about five iterations.  Thus by consistency of
iteration the angular distribution eventually determines
 the energy spectrum. The energy spectrum obtained this way
may serve as a method to measure the incident
energy spectrum.

If one assumes that the density profile of the Earth is
already well
known, then the angular distribution strongly overdetermines the problem.
This is quite interesting. One might even be able to deduce the
 energy dependence of the neutrino-nucleon total cross section. There is
 an opportunity, then, to explore fundamental features of particle
 physics which
have never been directly measured. While the predictions of small-x 
QCD are thought to be reliable in this region, there are various reasons
 (Ralston {\it et al} 1996b) why direct verification is desirable, and 
 might yield new physical information.

\section {Results and Discussion:}
\label{results.sec}
  One boundary
condition must be fixed in order to obtain
convergence and a unique solution.  The boundary condition
 is taken to be the value of
 the density of the Earth near the
surface. Since the surface density is known with reasonable accuracy, the
boundary
condition is not a cause of substantial uncertainty. 
The simulations used the Preliminary 
Earth Reference Model (PREM) for the Earth's density profile
 (see Jain, Ralston and Frichter 1999 for details). Two moments of
 the density are
well known: These are the total mass of the Earth (just as Cavendish used),
and the Earth's moment of inertia.  But the entire procedure can be
carried out (as we did) without making use of these moments, so in
 practice the density we find
is over-determined.

We employed a generic form of the diffuse AGN neutrino flux,
$\Phi_\nu(E)=\Phi_oE^{-2}$ for $10\ {\rm TeV} < E < 10^4\ {\rm TeV}$. This
form is within the range of current theoretical predictions. Neutrino
energies in the simulations (done by Monte Carlo) 
 varied from $10 - 10^4$ TeV. The practical lower limit of energy is between 10 and 50
TeV, since below this value the Earth is essentially
transparent. The practical upper limit is between $10^3$ and $10^4$ TeV, beyond
which the Earth is essentially opaque, and fluxes are
 expected to be too small
for tomography.

We assumed 1000 receivers operating for 2 years, or 200 receivers operating
 for 10 years. The error bars in the plots reflect the statistical errors
in density determination for different flux assumptions. 
 To a good approximation the statistical errors scale with the number of
 receivers, so that 10 times fewer receivers with the same flux is equivalent
to the assumed number of receivers with 10 times smaller flux.
 The final extracted density profiles for two cases are shown in
Figs. (1,2).  The step between the core and lower mantle
 is very well resolved in Fig. (1).  However the inner core is not
 resolved, partly
 because we did not make a special effort to bin events in a way that
 might bring out this feature. The inner core is difficult to resolve in
 any event, because it subtends a rather small solid angle.
 The total mass is consistent, so that
 Neutrinos can ``weigh the Earth". 
On the other hand, since the mass is well known, then by iteration the 
boundary condition at the ``crust'' can be deduced.

There are certainly many uncertainties in going from our model
calculations to realistic
experimental design. Only further study can resolve this.  During a period
of a few years to a decade, a $KM3$ neutrino telescope will be fulfilling a
primary mission as a fundamentally new kind of instrument for observing the
cosmos. With the same kind of detector, neutrino tomography
could also provide important and independent information about the Earth's
interior. Further details are given in Jain, Ralston and Frichter (1999).

\section {Acknowledgements:}
Supported
by DOE grant numbers DE-FGO2-98ER41079, DAE/PHY/96152, 
the KU General Research Fund and NSF-K*STAR
Program under the Kansas Institute for Theoretical and Computational
Science. This paper was written while P. Jain and J. Ralston were visiting
ICTP, Trieste. They thank the ICTP staff for hospitality.

%
\vspace{1ex}
\begin{center}
{\Large\bf References}
\end{center}
%
Frichter, G. M.,  Ralston, J. P. \&  McKay, D. W. 1995, Phys. Rev.
Lett. 74, 1508-1511\\
Frichter, G. M.,  Ralston, J. P. \& McKay, D. W. 1996a, Phys.
Rev. D53
1684\\
Frichter, G. M.,  Ralston, J. P. \& McKay, D. W. 1996b, astro-ph 9606007,
in {\it Proceedings of the Seventh International Symposium on 
Neutrino Telescopes} (Venice 1996), M. Baldo-Ceolin, editor\\
Gandhi, R., Quigg, C., Reno, M. H. \& Sarcevic, I. 1998, Phys. Rev. D
58  093009.\\
Jain, P., Ralston, J.P., \& Frichter, G.M., 1999, hep-ph 9902206, to
 appear in {\it Astroparticle Physics}\\
Kuo, C.,  Crawford, H. J., Jeanloz, R., Romanowicz, B., 
Shapiro G. \&  Stevenson, M. L. 1995, Earth and Planetary Science Letters {\bf 133},
95, and references therein\\
Ralston, J. P., McKay, D. W. \& Frichter, G.M. 1996, astro-ph 9606008,
in {\it Proceedings of the Seventh International Symposium on 
Neutrino Telescopes} (Venice 1996), M. Baldo-Ceolin, editor\\
Stecker, F. W., Done, C., Salamon M. H. \& Sommers, P. 1992, Phys.
Rev. Lett. {\bf 66}, 2697; E (1992): {\it ibid} {\bf 69}, 2738\\  
Szabo A. P. \&  Protheroe, R. J. 1994, Astropart. Phys. {\bf 2},
375\\ 
Protheroe R. J. \&  Szabo, A. P. 1992, Phys. Rev. Lett. {\bf 69}, 2285\\

\begin{figure} [t,b] \hbox{\hspace{4em}
\hbox{\epsfig{file=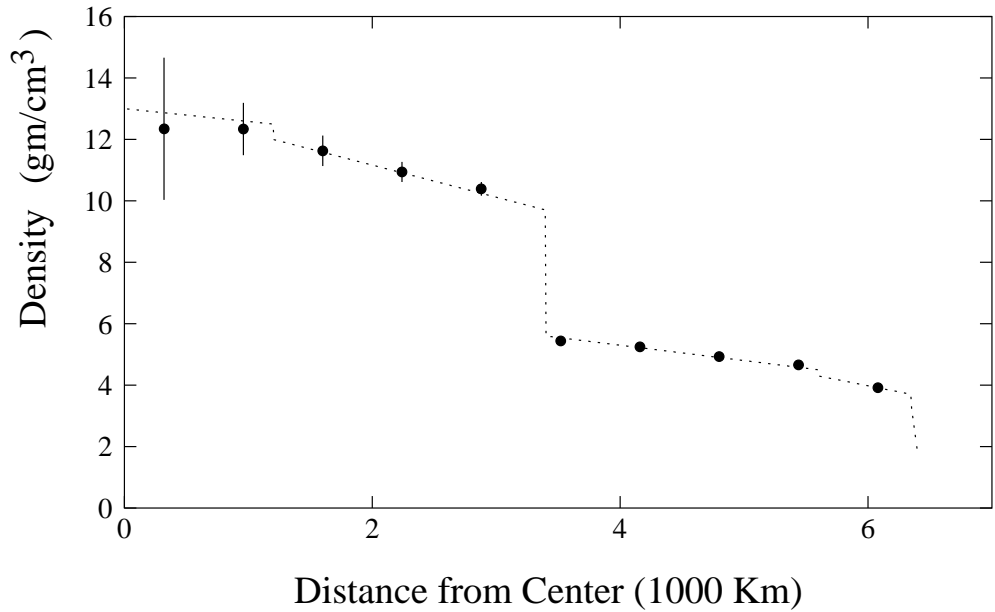,height=8cm}}}
\bigskip
\caption{The extracted nucleon density using ten radial bins along with the
statistical errors. The PREM model for the Earth's density, dashed curve,
was used to generate data. The step between the core and lower mantle is
very well resolved, while the difference between the inner and outer cores
is not.} \label{fig4} \end{figure}

\begin{figure} [t,b] \hbox{\hspace{4em}
\hbox{\epsfig{file=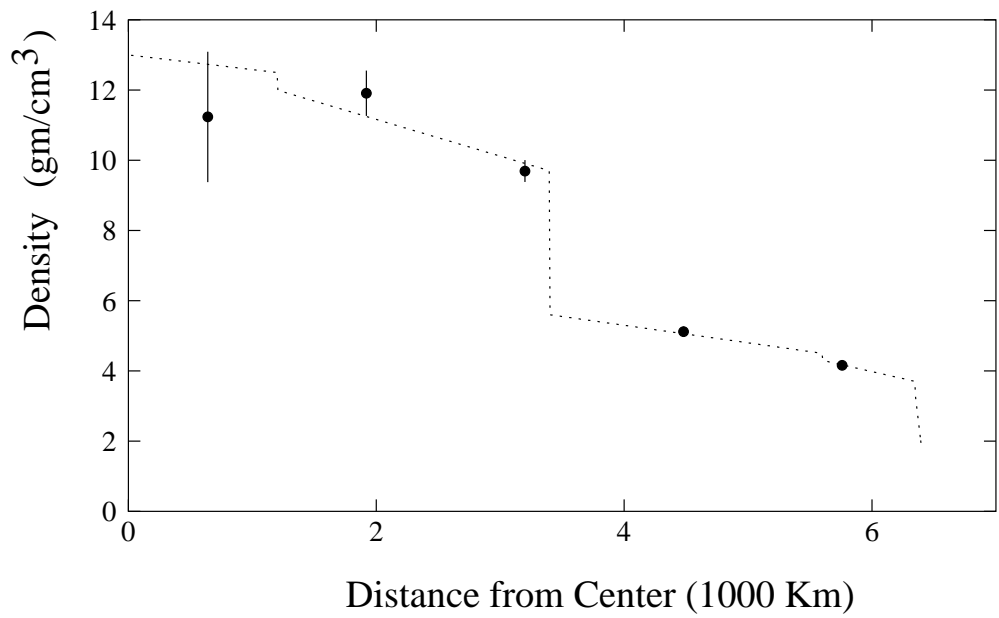,height=8cm}}}
\bigskip
\caption{The extracted density using five radial bins along with the
statistical errors, assuming one tenth the event rate compared to Fig. 1.
The step between the core and lower mantle remains well resolved. }
\label{fig5} \end{figure}
\end{document}